\documentclass[aps,prd,nofootinbib]{revtex4}
\usepackage{graphicx}
\usepackage{rotating}

\begin{document}
\title{Multiple inflation and the WMAP `glitches'} 
\author{Paul Hunt and Subir Sarkar}
\medskip
\affiliation{Theoretical Physics, University of Oxford, 1 Keble Road, 
             Oxford OX1 3NP, UK}
%\date{\today}

\begin{abstract}
Observations of anisotropies in the cosmic microwave background by the
{\sl Wilkinson Microwave Anisotropy Probe} suggest the possibility of
oscillations in the primordial curvature perturbation. Such deviations
from the usually assumed scale-free spectrum were predicted in the
multiple inflation model wherein `flat direction' fields undergo rapid
phase transitions due to the breaking of supersymmetry by the large
vacuum energy driving inflation. This causes sudden changes in the
mass of the (gravitationally coupled) inflaton and interrupts its slow
roll. We calculate analytically the resulting modifications to the
curvature perturbation and demonstrate how the oscillations arise.
\end{abstract}

\pacs{98.80.Cq, 98.70.Vc}
\maketitle

\section{Introduction}

Despite the overall success of the standard $\Lambda$CDM inflationary
model in matching the results from WMAP, it is striking that the
goodness-of-fit statistics for the data are rather poor
\cite{Spergel:2003cb}. In particular, $\chi_{\rm eff}^2/\nu =
1432/1342$ for the fit to the TT spectrum which implies a probability
of only $\sim3\%$ for the best fit model to be correct. The lack of
power at large scales relative to the standard $\Lambda$CDM model
prediction has motivated several alternative proposals for the
spectrum and nature of the primordial fluctuations
\cite{Bridle:2003sa,Contaldi:2003zv,Blanchard:2003du,Cline:2003ve,Feng:2003zu,Kawasaki:2003dd,
Bastero-Gil:2003bv,Kaloper:2003nv,Tsujikawa:2003gh,Moroi:2003pq,Piao:2003wp},
however the uncertainties (and cosmic variance) at low multipoles are
large and it has been argued that the observed low quadrupole (and
octupole) are unlikely only at about a few per cent level
\cite{Gaztanaga:2003ub,Efstathiou:2003wr,deOliveira-Costa:2003pu}.\footnote{However
there is an unexpected alignment between the quadrupole and octupole
\cite{deOliveira-Costa:2003pu,Schwarz:2004gk} and there appear to be
systematic differences between the north and south ecliptic
hemispheres \cite{Eriksen:2003db}, so this issue cannot be considered
settled as yet.} In fact attempts to judge the goodness-of-fit by eye
can be misleading since neighbouring $C_l$'s are correlated (being
`pseudo-$C_l$'s evaluated on a `cut' sky); taking this into account it
is found \cite{Spergel:2003cb} that the excess $\chi^2$ comes mainly
from sharp features or `glitches' in the power spectrum that the model
is unable to fit. In particular there are glitches at $l\sim120$, 200
and 340 amongst the first and second acoustic peaks.

The WMAP team have noted that these glitches may have been caused by
beam asymmetry, gravitational lensing of the CMB, non-Gaussianity in
the noise maps, or the method of power spectrum reconstruction from
the CMB maps \cite{Spergel:2003cb}. However they have also considered
the possibility that these glitches are due to features in the
underlying primordial curvature perturbation spectrum
\cite{Peiris:2003ff}, specifically in the `multiple inflation' model
\cite{Adams:1997de} where sudden changes in the mass of the inflaton
can generate characteristic localized oscillations in the
spectrum. This is physically well motivated since spontaneous symmetry
breaking can occur {\em during} inflation for `flat directions' in
supersymmetric theories and such fields are gravitationally coupled to
the inflaton \cite{Adams:1997de}. The resulting change in the
potential of the inflaton field $\phi$ was modeled earlier as
\cite{Adams:2001vc}
\begin{equation}
 V (\phi) = \frac{1}{2} m_{\phi}^2 \phi^2 
  \left[1 + c_{\rm ampl} \tanh 
  \left(\frac{\phi - \phi_{\rm step}}{d_{\rm grad}}\right)\right],
\label{tanh}
\end{equation}
which describes a standard `chaotic inflation' potential with a step
starting at $\phi_{\rm step}$ with amplitude and gradient determined
by the parameters $c_{\rm ampl}$ and $d_{\rm grad}$ respectively. This
was shown to result in oscillations in the primordial curvature
perturbation spectrum by numerically solving the governing
Klein-Gordon equation \cite{Adams:2001vc}. The WMAP team found that
the fit to the data improves if such oscillations are allowed for,
with a reduction of $\chi^2$ by 10 for the model parameters $\phi_{\rm
step}=15.5\,M_{\rm P}$, $c_{\rm ampl} = 9.1 \times 10^{-4}$ and
$d_{\rm grad} = 1.4 \times 10^{-2} M_{\rm P}$ \cite{Peiris:2003ff}.

We calculate the curvature perturbation spectrum from multiple
inflation more realistically, taking into account the evolution of
both fields as dictated by the inflationary dynamics
\cite{Adams:1997de}, and using a WKB approximation
\cite{Martin:2002vn}, rather than numerical integration to solve the
governing equation, in order to gain insight into how such
oscillations are generated. In a subsequent publication we perform
fits to the WMAP data with the cosmological parameters {\em
unconstrained}, in order to investigate the sensitivity of their
inferred values to such deviations from a scale-free spectrum
\cite{paper2}.

The possibility that the WMAP glitches are due to oscillations in the
primordial spectrum has also been considered in the context of
inflationary models invoking `trans-Planckian' physics
\cite{Wang:2002hf,Burgess:2002ub,Martin:2003sg,Okamoto:2003wk,Martin:2004iv}.
Sharp features in the primordial spectrum may also have been generated
by resonant particle production
\cite{Chung:1999ve,Elgaroy:2003hp,Kawasaki:2004pi}.  Several authors
have attempted to reconstruct the curvature perturbation from the WMAP
data and have noted possible features in its spectrum
\cite{Bridle:2003sa,Mukherjee:2003ag,Hannestad:2003zs,Shafieloo:2003gf,Kogo:2003yb,
Tocchini-Valentini:2004ht}. It is clearly necessary to have an
analytic formulation of the modification to the usually assumed
scale-free spectrum of the inflationary curvature perturbation, both
in order to provide a link between CMB observables and the physics
responsible for the glitches, and to distinguish between the various
suggestions for such new physics.

\section{Multiple Inflation}

Supergravity (SUGRA) theories consist of a `visible sector' and a
`hidden sector' coupled together gravitationally
\cite{Nilles:1983ge,Chung:2003fi}. The Standard Model particles are
contained in the visible sector and supersymmetry (SUSY) breaking
occurs in the hidden sector. Most SUGRA models of inflation have the
inflaton situated in the hidden sector. This is because it is easier
to protect the necessary flatness of the inflaton potential against
radiative corrections there. The visible sector is usually assumed to
be unimportant during inflation.

However this might not be true if one of the visible sector fields
undergoes gauge symmetry breaking. Suitable candidates for this are
the so-called flat direction fields which generically exist in
supersymmetric theories (for a review, see
ref.\cite{Enqvist:2003gh}). These are directions in field space in
which the potential vanishes, in the limit of unbroken SUSY. During
inflation in supergravity theories scalar fields typically receive a
contribution of $m^2 \sim \pm \mathcal{O}(H^2)$ to their mass-squared
due to SUSY breaking by the large vacuum energy driving inflation
\cite{Coughlan:yk,Copeland:1994vg,Dine:1995uk}. (For the inflaton this
constitutes the notorious `$\eta$ problem' \cite{Lyth:1998xn}; in
common with most models of inflation we need to assume that some
mechanism reduces the inflaton mass-squared by a factor of a least 20
in order to allow sufficient inflation to occur.) The SUSY-breaking
mass term typically dominates the potential along the flat
directions. If the mass-squared is negative at the origin in field
space, the flat direction $\rho$ will be stabilized at an intermediate
scale vev of
\begin{equation}
\Sigma \sim \left(M_{\rm P}^{n-4}m^2\right)^{1/\left(n-2\right)},
\end{equation} 
by higher dimensional operators $\propto \rho^n/M_{\rm P}^{n-4}$
\cite{Adams:1997de}. These appear in the potential after integrating
out heavy degrees of freedom, because we are working in an effective
field theory valid below some cut-off scale, which we take to be the
reduced Planck mass: $M_{\rm P} \equiv (8\pi\,G_{\rm N})^{-1/2} \simeq
2.4 \times 10^{18}$ GeV.

We assume that the inflationary era when the observed curvature
perturbation was produced was preceeded by a hot phase, so that all
fields were initially in thermal equilibrium. The potential along the
flat direction is then \cite{Adams:1997de}
\begin{eqnarray}
V (\rho, T) & \simeq & \left\{
\begin{array}{ll}
C_1 T^2 \rho^2, & \mathrm{for} \quad \rho\ll T,\\
-m^2 \rho^2 + \frac{1}{90} \pi^2 N_{\rm h} (T) T^4 
 + \frac{\gamma \rho^n}{M_{\rm P}^{n-4}}, & \mathrm{for} \quad T \ll \rho < \Sigma,
\end{array}\right.
\end{eqnarray}
with a smooth interpolation at $\rho \sim T$. Here we have included
the 1-loop finite temperature correction to the potential
\cite{Yamamoto:1985rd,Barreiro:1996dx} with $N_{\rm h} (T)$ being the
number of helicity states with mass much less than the temperature.

\begin{figure}[tbh]
\includegraphics{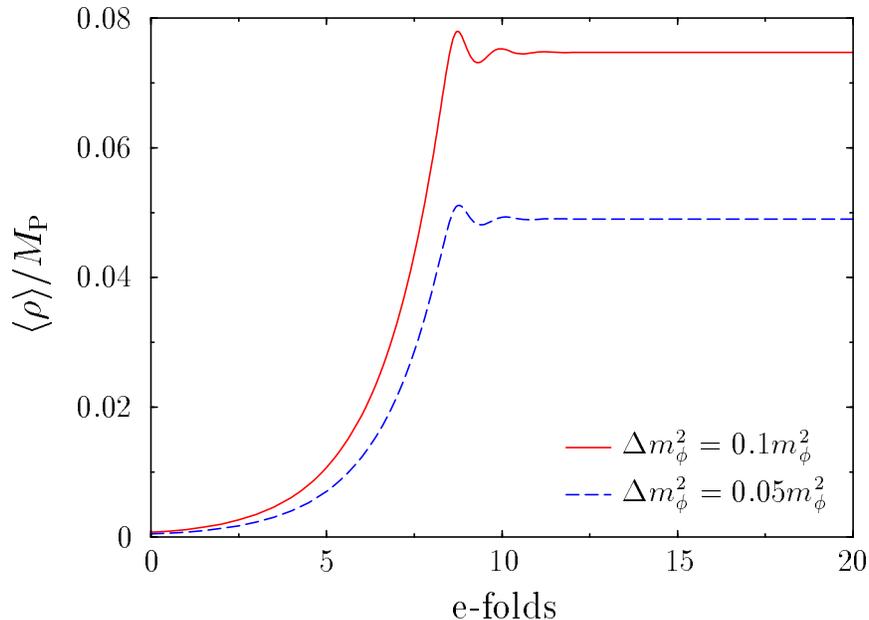}
\caption{\label{ev} The evolution of the flat direction field $\rho$,
 corresponding to two different values for the change in the inflaton
 mass.}
\end{figure}

The finite temperature correction therefore creates a barrier of
height $\mathcal{O}(T^4)$ in between the origin and $\rho \sim T^2/m$.
The tunneling rate through the barrier is negligible and so $\rho$ is
confined at the origin until the temperature, which falls rapidly
during inflation, reaches $T \sim m$ and the barrier disappears. The
flat direction field $\rho$ then undergoes a phase transition as it
evolves to its global minimum at $\Sigma$. This is governed by the
usual equation of motion
\begin{equation}
\ddot{\rho} + 3H \dot{\rho} = -\frac{dV}{d\rho},
\end{equation} 
which has solutions
\begin{eqnarray}
\rho & \simeq & \left\{\begin{array}{ll}
\rho_0 \exp \left[\frac{3Ht}{2}
\left(\sqrt{1 + \frac{8m^2}{9H^2}}-1\right)\right], 
& \left\langle\rho\right\rangle \ll\Sigma,\\
\\ \Sigma + K_1 \exp \left(-\frac{3Ht}{2}\right) \sin\left[\frac{3Ht}{2}
\sqrt{(n-2)\frac{8m^2}{9H^2}-1} + K_2\right], & 
\left\langle \rho\right\rangle \sim\Sigma.\end{array}\right.
\label{pt}
\end{eqnarray}
Here we have taken
\begin{eqnarray}
\frac{\mathrm{d}V}{\mathrm{d}\rho} & \simeq & \left\{ \begin{array}{ll}
-2m^{2}\rho, & \left\langle \rho\right\rangle \ll\Sigma,\\
\left(\rho-\Sigma\right)\left.\frac{\mathrm{d}^2V}{\mathrm{d}\rho^{2}}
\right|_{\rho=\Sigma},
 & \left\langle \rho\right\rangle \sim\Sigma.\end{array}\right.
 \end{eqnarray}
Therefore $\rho$ evolves exponentially towards its minimum, with most
of the growth occurring over the last $\sim 2$ e-folds
\cite{Adams:1997de}, and then performs a few strongly damped
oscillations about $\Sigma$, as seen in Fig.\,\ref{ev}.
(We assume that damping by particle creation as $\rho$ oscillates
about its minimum is less important than the severe damping due to the
large inflationary Hubble parameter evident in Fig.\,\ref{ev}.)

After the phase transition the vacuum energy density driving inflation
is reduced by a factor of $\left[1 - (\Sigma/M_{\rm P})^2\right]$. For
$\Sigma$ significantly below the Planck scale this can be neglected,
hence $H$ can be taken to be sensibly constant. A more important
effect occurs due to the {\em gravitational} coupling between $\rho$
and the inflaton field $\phi$.

The particular background inflationary cosmology in which the phase
transition occurs is not directly relevant for our work. For
definiteness we take it to be a `new inflation' potential with a cubic
leading term which is a phenomenologically successful model with a
naturally small inflaton mass \cite{Ross:1995dq,Adams:1996yd}. Then,
for example, a term $\kappa\phi\phi^\dagger\rho^2/M_{\rm P}^2$ in the
K\"{a}hler potential which is allowed by symmetry (near $\phi \sim 0$)
yields a coupling term $\lambda \phi^2 \rho^2/2$ in the inflaton
potential, where $\lambda = \kappa m^2/M_{\rm P}$
\cite{Adams:1997de}. With
\begin{equation}
 V (\phi, \rho) = V_{0} -  c_3 \phi^3 
                  + \frac{1}{2}\lambda\phi^2\rho^2+ \ldots,
\end{equation}
this causes the effective mass-squared $m_\phi^2 \equiv {\rm d}^2
V/{\rm d}\phi^2$ of the inflaton to change from $m_\phi^2 = -6 c_3
\langle\phi\rangle$ before the phase transition, to $m_\phi^2 = -6 c_3
\langle\phi\rangle + \lambda\Sigma^2$ afterwards. This will affect
the curvature perturbation spectrum which is very sensitive to the
mass of the inflaton.

In order to produce observable effects in the CMB or large-scale
structure the phase transition must take place as cosmologically
relevant scales are leaving the horizon. In supersymmetric theories
there are many flat directions which can potentially undergo symmetry
breaking during inflation, so it is not unlikely that a phase
transition occurred in the $\sim8$ e-folds which is sampled by
observations \cite{Adams:1997de}, particularly since the observations
suggest that (the last period of) inflation may not have lasted much
longer than the minimum necessary to generate an universe as big as
the present Hubble volume.

\section{The Curvature Perturbation}

Some care is required in calculating the curvature perturbation
spectrum for multiple inflation because we might {\em a priori} expect
two-field inflation during the phase transition, when $\phi$ and
$\rho$ evolve simultaneously. This might mean that both curvature as
well as isocurvature perturbations are created, involving a
significantly more complicated computation
\cite{Garcia-Bellido:1995qq}. However, as we show below, our model of
multiple inflation is effectively single-field inflation as far as the
generation of curvature perturbations is concerned.

The curvature perturbation on comoving hypersurfaces $\mathcal{R}$ is
given by $\mathcal{R=\delta}\mathcal{N}$, where $\mathcal{N}$ is the
number of e-folds of expansion measured locally by a comoving observer
passing from an initial, spatially flat, slice of space-time to a
final, comoving, slice \cite{Sasaki:1995aw}. The perturbations
$\delta\phi_{\rm fts}$ and $\delta\rho_{\rm fts}$ are defined on the
spatially flat time slice and the final comoving slice defines
$\mathcal{R}$. The first time slice is taken during inflation, soon
after horizon crossing, and the second time slice is taken at the end
of inflation, after $\mathcal{R}$ has become constant. During the
phase transition the slow-roll conditions in $\phi$ are no longer
satisfied but we assume the end of inflation to be well after this
epoch. Then we have
\begin{equation}
 \mathcal{R} = \frac{\partial \mathcal{N}}{\partial\phi} \delta\phi_{\rm fts} + 
               \frac{\partial \mathcal{N}}{\partial\rho} \delta\rho_{\rm fts},
\end{equation}
and so
\begin{equation}
 \mathcal{P_R} = \left(\frac{\partial \mathcal{N}}{\partial\phi}\right)^2 
 \mathcal{P}_{\phi} + \left(\frac{\partial \mathcal{N}}{\partial\rho}\right)^2 
 \mathcal{P}_{\rho} \simeq 
 \left(\frac{{\rm d}\mathcal{N}}{{\rm d}\phi}\right)^2 \mathcal{P}_{\phi},
\end{equation}
which is just the single-field inflation expression for
$\mathcal{P_R}$. This is because the vacuum energy is {\em dominated}
by $\phi$, which drives inflation, and $\rho$ has a negligible effect
on the number of e-folds of inflation.

The curvature perturbation spectrum is usually calculated analytically
using the slow-roll approximation to some chosen order in the
slow-roll parameters \cite{Lyth:1998xn}. This assumes the potential is
relatively smooth, which is however inappropriate for multiple
inflation. For example, for the standard chaotic inflation potential
$V=\frac{1}{2}m_{\phi}^2\phi^2$, the usual slow-roll expression gives
\begin{equation}
 \mathcal{P_R}^{1/2} = \frac{1}{2\sqrt{3}\pi M_{\rm P}^3}
 \left|\frac{V^{3/2}}{{\rm d}V/{\rm d\phi}}\right| \simeq 
 \frac{1}{2\sqrt{3}\pi M_{\rm P}^3}
 \frac{V^{3/2}}{m_{\phi}^2\phi },
\end{equation}
where all quantities are evaluated at horizon crossing. This gives a
simple `step' in the amplitude if the inflaton mass changes suddenly
as in eq.(\ref{tanh}), missing all the fine detail that we expect in
the spectrum. (A discussion of the slow-roll approximation applied to
a potential with a sudden change in its slope \cite{Starobinsky:ts}
rather than in its curvature can be found in ref.\cite{Leach:2001zf}.)

Therefore we must resort to first principles and use the general
formalism \cite{Stewart:1993bc} for calculating the curvature
perturbation spectrum. Instead of the slow-roll approximation we use
the recently suggested WKB approximation
\cite{Martin:2002vn,Martin:2003bt}.

The metric describing scalar perturbations in a flat universe can be
parameterized as
\begin{equation} 
\mathrm{d}s^2=a^2\left[ \left(1+2A_\mathrm{s}\right)\mathrm{d}\eta^2-
2\partial_i B_\mathrm{s}\mathrm{d}\eta\mathrm{d}x^i-\left\{\left(1
-2D_\mathrm{s}\right)\delta_{ij}+2\partial_i\partial_jE_\mathrm{s}\right\}
\mathrm{d}x^i\mathrm{d}x^j\right].
\end{equation} 
We employ the gauge-invariant quantity
\begin{equation} 
u=a\delta\phi^{\left({\rm gi}\right)}+z\Psi=z\mathcal{R},
\end{equation} 
to characterize the perturbations; see
refs.\cite{Mukhanov:1990me,Malik:2001rm,Riotto:2002yw} for more details.
Here
\begin{eqnarray} 
\delta\phi^{\left({\rm gi}\right)} & = &
\delta\phi+\phi'\left(B_\mathrm{s}-E_\mathrm{s}'\right),\\ \textrm{$\Psi$} & = &
D_\mathrm{s}-\frac{a'}{a}\left(B_\mathrm{s}-E_\mathrm{s}'\right),
\end{eqnarray} 
are also
gauge-invariant, $z=a\dot{\phi}/H$, and 
\begin{equation} 
\mathcal{R}=D_\mathrm{s}+H\frac{\delta\phi}{\dot{\phi}}.
\end{equation}
The primes indicate derivatives with respect to conformal time
\begin{equation}
 \eta = \int\frac{{\rm d}t}{a} = -\frac{1}{aH},
\end{equation}
where the last equality holds in de Sitter space. The Fourier
components of $u$ satisfy the Klein-Gordon equation of motion
\begin{equation}
 u''_{k} + \left(k^2 - \frac{z''}{z}\right) u_k = 0 .
\label{kg}
\end{equation}
 The spectrum depends directly on $u_k$ and is given by
\begin{equation}
 \mathcal{P_R}^{1/2} = \sqrt{\frac{k^3}{2\pi^2}} \left|\mathcal{R}_k\right| 
 = \sqrt{\frac{k^3}{2\pi^2}}\left|\frac{u_k}{z}\right|.
\label{PR}
\end{equation}
The solutions of eq.(\ref{kg}) are governed by the relative sizes of
$k^2$ and $z''/z$. In de Sitter space $z''/z=2/\eta^2$, therefore
initially when $k^2 \gg z''/z$ and the mode $u_k$ is well inside the
horizon, we have the flat spacetime solution
\begin{equation}
 u_k = \frac{1}{\sqrt{2k}}{\rm e}^{-ik(\eta-\eta_i)},
\label{inside}
\end{equation}
where $\eta_{\rm i}$ is some arbitrary time at the beginning of
inflation.  In the opposite limit, $k^2 \ll z''/z$, we have the
solution
\begin{equation} 
 u_{k}=A_{k}z+B_{k}z\int^{\eta}\frac{{\rm d}\mu}{z^{2}\left(\mu\right)},
\label{outside}
\end{equation} 
when $u_{k}$ is well outside the horizon. The first term on the right
is the growing mode solution and the second is the decaying mode. Thus
$\mathcal{P_{R}}$ at late times is equal to
$k^{3}\left|A_{k}\right|^{2}/2\pi^{2}$.

We will use the WKB approximation to solve the mode equation, with
eq.(\ref{inside}) as an initial condition. The WKB approximation
cannot be directly applied to eq.(\ref{kg}) because the approximation
breaks down on super-horizon scales. However, as noted in
ref.\cite{Martin:2002vn}, the WKB approximation becomes relevant if we
switch to the variables
\begin{eqnarray}
 x & = & \ln\left(\frac{Ha}{k}\right), \\
 U & = & {\rm e}^{x/2} u_k.
\end{eqnarray}
The transformed mode equation is then
\begin{equation}
 \frac{{\rm d}^2 U}{{\rm d} x^2} + Q (x) U = 0, 
\label{mode}
\end{equation}
where the effective frequency,
\begin{equation}
 Q(x) = \left(1 - \frac{z''}{zk^2}\right){\rm e}^{-2x} - \frac{1}{4},
\label{freq}
\end{equation}
generally decreases with time during inflation, passing through zero
at least once.\footnote{$Q$ is related to the variable $g_s$ of
Refs.\cite{Habib:2002yi,Habib:2004kc} by $Q=-g_s\eta^2.$} In de Sitter
space, this is
\begin{equation}
 Q(x) = \left(\frac{k}{aH}\right)^2 - \frac{9}{4}.
\end{equation}
In the WKB formalism \cite{wkb} we expand $U$ as an asymptotic series
\begin{equation} 
U \left(x\right) = \exp\left[{\displaystyle
 \sum_{n=0}^{\infty}S_{n}\left(x\right)}\right], \quad
S_{n} \left(x\right) \gg S_{n+1} \left(x\right).
\label{zsx}
\end{equation}
Substituting this into eq.(\ref{mode}) results in the following series
of equations
\begin{eqnarray}
{S'_0}^2 & = & -Q, \\
2S'_0 S'_1 + S''_0 & = & 0, \\
2S'_0 S'_n + S''_{n-1} + \sum_{j=1}^{n-1} S'_j S'_{n-j} & = & 0, 
\quad {\rm for}\ n \geq 2,
\end{eqnarray}
where the primes indicate derivatives with respect to $x$.

This gives the first few terms as
\begin{eqnarray}
 & Q>0 & \left\{ \begin{array}{ccl}
S_{0} & = & \pm i\int^{x}Q^{1/2}{\rm d}y,\\
\\S_{1} & = & -\frac{1}{4}\ln Q,\\
\\S_{2} & = & \pm i\int^{x}\left(-\frac{Q''}{8Q^{3/2}}
+\frac{5Q^{2}}{32Q^{3/2}}\right){\rm d}y,\\
\\S_{3} & = & \frac{Q''}{16Q^{2}}
-\frac{5Q^{2}}{64Q^{3}},\end{array}\right.\\
 &  & \nonumber\\
 & Q<0 & \left\{ \begin{array}{ccl}
S_{0} & = & \pm\int^{x}\left(-Q\right)^{1/2}{\rm d}y,\\
\\S_{1} & = & -\frac{1}{4}\ln\left(-Q\right),\\
\\S_{2} & = & \pm\int^{x}\left[-\frac{Q''}{8\left(-Q\right)^{3/2}}
+\frac{5Q^{2}}{32\left(-Q\right)^{3/2}}\right]{\rm d}y,\\
\\S_{3} & = & \frac{Q''}{16Q^{2}}-\frac{5Q^{2}}{64Q^{3}}.
\end{array}\right.
\end{eqnarray}
Using only the first two terms in the expansion and neglecting the
rest gives the 1st-order WKB approximation (for $Q>0$)
\begin{eqnarray}
U_{\rm I} & = &
\frac{A}{Q^{1/4}(x)} \exp\left[i\int_{x_{\rm i}}^{x} Q^{1/2}(y) {\rm d}y \right] 
+ \frac{B}{Q^{1/4}(x)} \exp\left[-i\int_{x_{\rm i}}^{x} Q^{1/2}(y) {\rm d}y\right],
\label{3}
\end{eqnarray}
where $A$ and $B$ are constants. At early times $Q\simeq{\rm e}^{-2x}$
which means that the WKB solution will match eq.(\ref{inside}) if
$A=0$ and $B=1/\sqrt{2k}$.

The solutions become inaccurate close to the roots (`turning points')
$x_*$ of $Q (x_*)=0$. To continue the solution through a turning point
we therefore need to match the WKB solutions valid to the right and
left of the turning point, with a solution valid in the neighbourhood
of the turning point. We patch the solutions together using asymptotic
matching.

We consider the case of a single turning point and divide the $x$-axis
into three regions. Region I is defined as the region where $x<x_*$,
$Q>0$ and the WKB approximation is valid. Region II is the
neighbourhood of $x_*$ where $Q$ can be approximated by its tangent at
$x_*$. Region III is where $x>x_*$, $Q<0$, and the WKB solution is
accurate. We assume there is an overlap between regions I and II, and
also between regions II and III.

In region II we have
\begin{equation}
Q \simeq -\alpha X, \quad 
X \equiv x-x_{*}, \quad
\alpha \equiv -\left.\frac{{\rm d}Q}{{\rm d}x}\right|_{x=x_*} > 0 .
\end{equation}
We rewrite the integration limits of $U_{\rm I}$ (\ref{3}) as
\begin{equation}
\int_{x_{\rm i}}^{x} Q^{1/2} {\rm d}y = \int_{x_{\rm i}}^{x_*} Q^{1/2} {\rm d}y 
 + \int_{x_*}^{x} Q^{1/2} {\rm d}y 
\equiv \Gamma + \int_{x_*}^{x} Q^{1/2} {\rm d}y .
\end{equation}
Then in the overlap between regions I and II, $U_{\rm I}$ becomes
\begin{equation}
U_{\rm I} \simeq \frac{\alpha^{-1/4}}{\sqrt{2k}} 
 \left(-X\right)^{-1/4}\exp\left[-i\left\{-\frac{2}{3}\left(-X\right)^{3/2}
+ \Gamma\right\}\right] .
\label{UI}
\end{equation}
With $Q\simeq-\alpha X$, the solution to eq.(\ref{mode}) is
\begin{equation}
U_{\rm II} \simeq
C \textrm{Ai} \left(\alpha^{1/3}X\right) 
+ D \textrm{Bi} \left(\alpha^{1/3}X\right) ,
\label{will2}
\end{equation}
where Ai and Bi are Airy functions of the first and second kinds and
$C$ and $D$ are constants. For $X\ll0$, $U_{\rm II}$ becomes
\begin{eqnarray} 
U_{\rm II} & \simeq &
\frac{\alpha^{-1/12}}{2\sqrt{\pi}} \left(-X\right)^{-1/4}
\left[\left(D - iC\right)
\exp\left\{i\left(\frac{2}{3}\left(-X\right)^{3/2} + 
\frac{\pi}{4}\right)\right\}\right. \nonumber \\
& &
\left.+\left(D+iC\right)\exp\left\{-i\left(\frac{2}{3}\left(-X\right)^{3/2}
+\frac{\pi}{4}\right)\right\}\right] .
\end{eqnarray} 
Requiring this to match $U_{\rm I}$ (\ref{UI}) in the overlap region
fixes the coefficients
\begin{equation} 
C = iD, \quad
D = \sqrt{\frac{\pi}{2k}} \alpha^{-1/6} 
    \exp\left[-i\left(\Gamma + \frac{\pi}{4}\right)\right].
\end{equation}
For $X\gg0$ we have
\begin{eqnarray} 
U_{\rm II} & \simeq & \frac{\alpha^{-1/12}}{\sqrt{\pi}}X^{-1/4}
\left[\frac{C}{2}\exp\left(-\frac{2}{3}\alpha^{1/2}X^{3/2}\right)
+D\exp\left(\frac{2}{3}\alpha^{1/2}X^{3/2}\right)\right] .
\label{UII}
\end{eqnarray}
In region III, the 1st-order WKB solution is 
\begin{eqnarray}
U_{\rm III} & = & \frac{F}{\left[-Q(x)\right]^{1/4}}
 \exp\left\{\int_{x_*}^{x}\left[-Q(y)\right]^{1/2}{\rm d}y\right\} 
\nonumber \\ & &
+\frac{G}{\left[-Q(x)\right]^{1/4}} 
 \exp\left\{-\int_{x_*}^{x}\left[-Q(y)\right]^{1/2}{\rm d}y\right\} .
\label{UIII1}
\end{eqnarray} 
When $Q\simeq-\alpha X$, this becomes
\begin{equation}
U_{\rm III} \simeq \alpha^{-1/4} X^{-1/4} 
\left[F\exp\left(\frac{2}{3}\alpha^{1/2}X^{3/2}\right)
+ G\exp\left(-\frac{2}{3}\alpha^{1/2}X^{3/2}\right)\right],
\end{equation}
which matches $U_{\rm II}$ (\ref{UII}) in the second overlap region if
\begin{equation}
F = \frac{\alpha^{1/6}}{\sqrt{\pi}}D,\quad
G = \frac{\alpha^{1/6}}{2\sqrt{\pi}}C.
\end{equation}
Substituting $U_{\rm III}$ into eq.(\ref{PR}) gives finally the curvature
perturbation spectrum
\begin{equation}
\mathcal{P_R^{\mathrm{1/2}}}\left( k \right) =
\left.\frac{H^2}{2\pi\dot{\phi}}
\left(\frac{k}{aH}\right)^{3/2}\right|_{x=x_{\rm f}} 
\left[-Q\left(x_{\rm f} \right)\right]^{-1/4}
\exp\left\{\int_{x_*}^{x_{\rm f}}\left[-Q(y)\right]^{1/2}{\rm d}y\right\}
\label{wkbex1} .
\end{equation}
where $x_{\rm f}$ is the value of $x$ at some late time well after
horizon crossing when $\mathcal{R}$ is constant, and we have neglected
the decaying mode proportional to $G$ which is negligible at late
times.

We now consider the 2nd-order WKB approximation where the first three
terms in the expansion (\ref{zsx}) are retained. At early times the
WKB solution is
\begin{eqnarray} 
U_{\rm I} & = &
\frac{1}{\sqrt{2k}Q^{1/4}}
\exp\left[-i\int_{x_{\rm i}}^x
\left(Q^{1/2} - \frac{Q''}{8Q^{3/2}} + 
\frac{5Q^{'2}}{32Q^{3/2}}\right){\rm d}y\right], \nonumber \\
& = &
\frac{1}{\sqrt{2k}Q^{1/4}}
\exp\left[-i\left(\int_{x_{\rm i}}^{x}
Q^{1/2}{\rm d}y - \frac{5Q'}{48Q^{3/2}} - 
\int_{x_{\rm i}}^{x} \frac{Q''}{48Q^{3/2}} {\rm d}y\right)\right],
\end{eqnarray}
where the second line follows after an integration by parts. As before
we rewrite the integration limits as
\begin{eqnarray}
\int_{x_{\rm i}}^{x} \left(Q^{1/2}{\rm d}y
 - \frac{Q''}{48Q^{3/2}}\right){\rm d}y
& = &
\int_{x_{\rm i}}^{x_*} Q^{1/2}{\rm d}y - \int_{x_{\rm i}}^{x_* -
\varepsilon}\frac{Q''}{48Q^{3/2}}{\rm d}y \nonumber \\
& &
+ \int_{x_*}^{x} Q^{1/2}{\rm d}y -
\int_{x_* - \varepsilon}^{x} \frac{Q''}{48Q^{3/2}}{\rm d}y , \nonumber\\
& \equiv &
\Upsilon + \int_{x_*}^{x} Q^{1/2}{\rm d}y
- \int_{x_* - \varepsilon}^{x}\frac{Q''}{48Q^{3/2}}{\rm d}y .
\end{eqnarray}
Here $\varepsilon$ is a small parameter introduced to regularize the
divergent integral $S_2$. This time, region II is the neighbourhood of
$x_*$ where
\begin{equation}
Q \simeq - \alpha X + \beta X^2, \quad 
\beta \equiv \left.\frac{1}{2}\frac{{\rm d}^2 Q}{{\rm d}x^2}\right|_{x=x_*} .
\label{expd}
\end{equation}
Then in the overlap of regions I and II
\begin{eqnarray} 
U_{\rm I} & \simeq &
\frac{\alpha^{-1/4}}{\sqrt{2k}}\left(-X\right)^{-1/4}
\left(1 + \frac{\beta X}{4\alpha}\right)
\exp\left[-i\left\{-\frac{2}{3}\left(-X\right)^{3/2}\right.\right. 
\nonumber \\ & &
-\frac{\beta}{5\alpha^{1/2}}\left(-X\right)^{5/2} 
+ \frac{5}{48\alpha^{1/2}}\left(-X\right)^{-3/2} 
- \frac{\beta}{12\alpha^{3/2}}\left(-X\right)^{-1/2} \nonumber \\
& &
\left.\left. 
+ \frac{\beta}{12\alpha^{3/2}}\varepsilon^{-1/2} + \Upsilon\right\}\right] .
\end{eqnarray} 
With $Q\simeq-\alpha X+\beta X^{2}$, eq.(\ref{mode}) has the
approximate solution
\begin{eqnarray}
U_{\rm II} & \simeq & K\left(1 + \frac{\beta X}{5\alpha}\right)
\textrm{Ai}\left[\alpha^{1/3}X\left(1-\frac{\beta X}{5\alpha}\right)\right] 
\nonumber \\
 &  & + L \left(1+\frac{\beta X}{5\alpha}\right)
\textrm{Bi}\left[\alpha^{1/3}X\left(1-\frac{\beta X}{5\alpha}\right)\right] .
\label{will}
\end{eqnarray}
For $X\ll0$, 
\begin{eqnarray}
U_{\rm II} & \simeq &
 \frac{\alpha^{-1/12}}{2\sqrt{\pi}}\left(-X\right)^{-1/4}
\left(1 + \frac{\beta X}{4\alpha}\right) \nonumber \\
 &  & \times\left[\left(L-iK\right) 
\exp\left\{i\left(\frac{2}{3}\left(-X\right)^{3/2} 
+ \frac{\beta}{5\alpha^{1/2}}\left(-X\right)^{5/2} 
+ \frac{\pi}{4}\right)\right\}\right. \nonumber \\
 &  &
\left. + \left(L+iK\right)
\exp\left\{-i\left(\frac{2}{3}\left(-X\right)^{3/2}
+ \frac{\beta}{5\alpha^{1/2}}\left(-X\right)^{5/2}
+ \frac{\pi}{4}\right)\right\}\right] ,
\end{eqnarray}
which matches $U_{\rm I}$ in the overlap region provided that
\begin{equation}
K = iL, \quad L =
\sqrt{\frac{\pi}{2k}}\alpha^{-1/6}\exp\left[-i\left(\Upsilon+\frac{\pi}{4}
-\frac{\beta}{12\alpha^{3/2}}\varepsilon^{-1/2}\right)\right] .
\end{equation}
When $X\gg0$, we have
\begin{eqnarray}
U_{\rm II} & \simeq & \frac{\alpha^{-1/12}}{\sqrt{\pi}}X^{-1/4} 
\left(1+\frac{\beta X}{4\alpha}\right)
\left[\frac{K}{2}\exp\left(-\frac{2}{3}\alpha^{1/2}X^{3/2}\right.\right.
\nonumber \\ &  &
\left.\left.+\frac{\beta}{5\alpha^{1/2}}X^{5/2}\right) 
+ L\exp\left(\frac{2}{3}\alpha^{1/2}X^{3/2} 
- \frac{\beta}{5\alpha^{1/2}}X^{5/2}\right)\right] .
\end{eqnarray}
The 2nd-order WKB solution in region III is 
\begin{eqnarray}
U_{\rm III} & = & \frac{M}{\left(-Q\right)^{1/4}}
\exp\left[ \int_{x_{*}}^{x}\left\{\left(-Q\right)^{1/2}{\rm d}y
- \frac{Q''}{8\left(-Q\right)^{3/2}}
+ \frac{5Q^{'2}}{32\left(-Q\right)^{3/2}}\right\}{\rm d}y\right]
\nonumber \\
 &  & +\frac{N}{\left(-Q\right)^{1/4}}
\exp\left[-\int_{x_{*}}^{x}\left\{\left(-Q\right)^{1/2}{\rm d}y
- \frac{Q''}{8\left(-Q\right)^{3/2}}
+ \frac{5Q^{'2}}{32\left(-Q\right)^{3/2}}\right\}{\rm d}y\right] ,
\nonumber \\
 & = & \frac{M}{\left(-Q\right)^{1/4}}
\exp\left[\int_{x_{*}}^{x}\left(-Q\right)^{1/2}{\rm d}y
- \frac{5Q'}{48\left(-Q\right)^{3/2}}
- \int_{x_*+\varepsilon}^{x}
\frac{Q''}{48\left(-Q\right)^{3/2}}{\rm d}y\right] \nonumber \\
 &  &
 +\frac{N}{\left(-Q\right)^{1/4}}
\exp\left[-\int_{x_{*}}^{x}\left(-Q\right)^{1/2}{\rm d}y
+ \frac{5Q'}{48\left(-Q\right)^{3/2}}
+ \int_{x_*+\varepsilon}^{x}\frac{Q''}{48\left(-Q\right)^{3/2}}
{\rm d}y\right] .
\end{eqnarray}
With $Q\simeq-\alpha X+\beta X^2$, this becomes
\begin{eqnarray}
U_{\rm III} & \simeq & \alpha^{-1/4}X^{-1/4}\left(1+\frac{\beta
 X}{4\alpha}\right)\left[M\exp\left(\frac{2}{3}\alpha^{1/2}X^{3/2}-
 \frac{\beta}{5\alpha^{1/2}}X^{5/2}\right.\right.\nonumber\\
 &  & \left.+\frac{5}{48\alpha^{1/2}}X^{-3/2}
 +\frac{\beta}{12\alpha^{3/2}}X^{-1/2}-
 \frac{\beta}{12\alpha^{3/2}}\varepsilon^{-1/2}\right)\nonumber\\
 &  & + N \exp\left(-\frac{2}{3}\alpha^{1/2}X^{3/2}+
 \frac{\beta}{5\alpha^{1/2}}X^{5/2}-
 \frac{5}{48\alpha^{1/2}}X^{-3/2}\right.\nonumber\\
 &  & \left.\left.- \frac{\beta}{12\alpha^{3/2}}X^{-1/2} +
 \frac{\beta}{12\alpha^{3/2}}\varepsilon^{-1/2}\right)\right].
\label{uf}
\end{eqnarray}
Hence if
\begin{eqnarray}
M & = &
\frac{\alpha^{1/6}}{\sqrt{\pi}}
\exp\left(\frac{\beta}{12\alpha^{3/2}}\varepsilon^{-1/2}\right)L ,
\nonumber \\
N & = &
\frac{\alpha^{1/6}}{2\sqrt{\pi}}
\exp\left(-\frac{\beta}{12\alpha^{3/2}}\varepsilon^{-1/2}\right)K ,
\label{vandw}
\end{eqnarray}
then $U_{\rm III}$ matches $U_{\rm II}$. Therefore in the 2nd-order
approximation
\begin{eqnarray}
\mathcal{P_R^{\mathrm{1/2}}}\left( k \right) & =
& \left.\frac{H^2}{2\pi\dot{\phi}}
\left(\frac{k}{aH}\right)^{3/2}\right|_{x=x_{\rm f}}
\exp\left(\frac{\beta}{12\alpha^{3/2}}\varepsilon^{-1/2}\right)
\left[-Q\left(x_{\rm f}\right)\right]^{-1/4} \nonumber \\
&  &
\times\exp\left[\int_{x_*}^{x_{\rm f}}
\left(-Q\right)^{1/2}{\rm d}y-\frac{5Q'}{48\left(-Q\right)^{3/2}}
- \int_{x_*+\varepsilon}^{x_{\rm f}}\frac{Q''}{48\left(-Q\right)^{3/2}}
{\rm d}y\right] .
\label{wkb2eq}
\end{eqnarray}
A weak dependence on the parameter $\varepsilon$ arises because of the
incomplete cancellation between
$\exp\left(\frac{\beta}{12\alpha^{3/2}}\varepsilon^{-1/2}\right)$ and
$\exp\left[\int_{x_*+\varepsilon}^{x_{\rm
f}}\frac{Q''}{48\left(-Q\right)^{3/2}}{\rm d}y\right]$ above. This is
because we have truncated the expansion of $Q$ at 2nd-order (see
eq.(\ref{expd})) for computational convenience.

\section{Results}

\begin{figure}[tbh]
\includegraphics{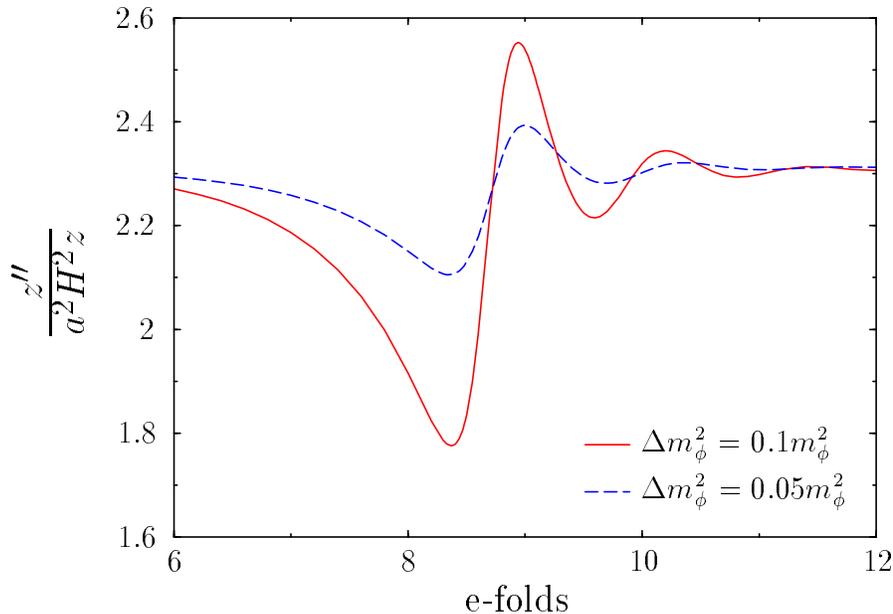}
\caption{ The evolution of $z''/z$ during the phase transition, for two
 different values of the change in the inflaton mass.}
\label{zzd} 
\end{figure}

During the phase transition, the scalar potential can be written as
\begin{equation}
V\left(\phi,\rho\right) = V_0 - c_3\phi^3 - m^2\rho^2 + \frac{1}{2}\lambda\phi^2\rho^2
 + \frac{\gamma}{M_{\rm  P}^{n-4}}\rho^n.
\end{equation}
Then the change in the inflaton mass-squared after the phase
transition is
\begin{equation} 
\Delta m_{\phi}^2 = \lambda\Sigma^2,\quad 
\Sigma =
\left[\frac{M_{\rm P}^{n-4}}{n\gamma}\left(2m^2
- \lambda\phi^2\right)\right]^{1/\left(n-2\right)}
\simeq \left(\frac{2m^2 M_{\rm P}^{n-4}}{n\gamma}\right)^{1/\left(n-2\right)}.
\end{equation}
The equations of motion are
\begin{eqnarray} 
\ddot{\phi}+3H\dot{\phi}=
& -\frac{\partial V}{\partial\phi} &
=\left(3c_3\phi-\lambda\rho^2\right)\phi, \label{eqm2} \\
\ddot{\rho}+3H\dot{\rho}= & -\frac{\partial V}{\partial\rho} &
=\left(2m^2-\lambda\phi^2-\frac{n\gamma}{M_{\rm P}^{n-4}}\rho^{n-2}\right)\rho.
\label{eqm}
\end{eqnarray}
Using these gives
\begin{equation}
\frac{z''}{z} = a^2\left(2H^2 + 6c_3\phi - \lambda\rho^2 
 - \frac{2\lambda\rho\dot{\rho}\phi}{\dot{\phi}}\right),
\end{equation}
which is shown in Fig.\,\ref{zzd}, taking $m^2_{\phi}\simeq0.05H^2$.
The effective frequency of eq.(\ref{freq}) is then obtained to be
\begin{equation}
Q = \frac{1}{H^2}\left(\frac{k^2}{a^2} - \frac{9}{4}H^2 - 6c_3\phi
 + \lambda\rho^2 + \frac{2\lambda\rho\dot{\rho}\phi}{\dot{\phi}}\right).
\end{equation}

\begin{figure}[tbh]
\includegraphics{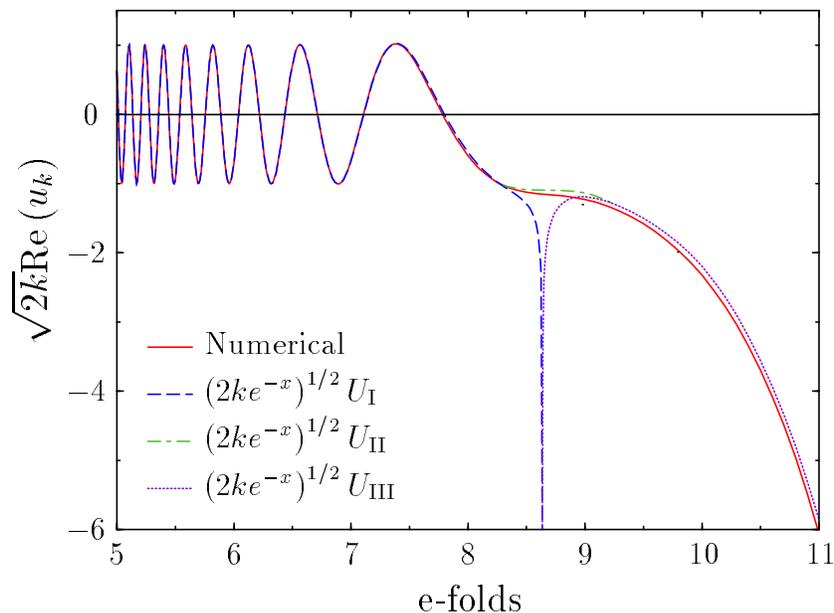}
\caption{ Comparison of the 1st-order WKB solution
(\protect\ref{will2}) with $u_k$ calculated numerically for $k =
6.7\times10^{-3}\; \rm {h\; Mpc^{-1}}$. The turning point is at
$x_*\simeq8.64$ and $\Delta m_\phi^2=0.1 m_\phi^2$. Note that the
solutions $U_{\rm I}$ (\protect\ref{3}) and $U_{\rm III}$
(\protect\ref{UIII1}) diverge at $x=x_*$ but $U_{\rm II}$
(\ref{will2}) is continuous and close to the exact solution.}
\label{mod} 
\end{figure}

\begin{figure}[tbh]
\includegraphics{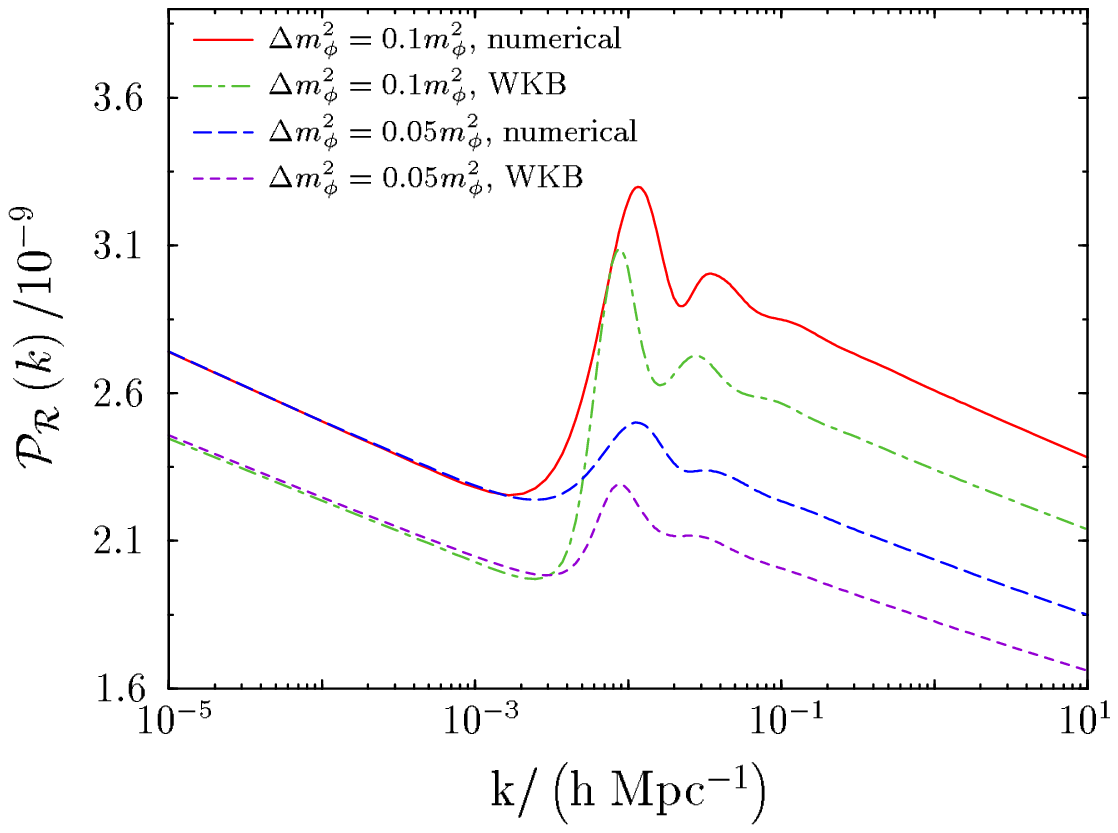}
\caption{ Comparison of the 1st-order WKB approximation with the
 numerically calculated exact spectrum, for two different values of the
 change in the inflaton mass.}
\label{biff} 
\end{figure}

We calculated the WKB solutions using $\phi\left(t\right)$ and
$\rho\left(t\right)$ found by numerical solution of eqs.(\ref{eqm2})
and (\ref{eqm}). An example of a 1st-order WKB solution is shown in
Fig.\,\ref{mod}, while Fig.\,\ref{biff} shows the power spectrum
obtained by numerically integrating eq.(\ref{kg}) mode by mode,
together with the 1st-order WKB approximation (\ref{wkbex1}), for two
different values of $\Delta m_\phi^2$. The spectrum $\mathcal{P_R}$
has the form of a positive step with superimposed damped oscillations
and a spectral index $n-1=\frac{d\ln\mathcal{P_{R}}}{d\ln k}$ of
$n\simeq0.961$. The 1st-order WKB spectra are too small by about 10\%
at large and small scales. They do exhibit oscillations but these are
slightly out of phase. We can understand the spectrum as follows.

The equation $Q=0$ can be rewritten as 
\begin{equation}
k = s (t), \quad
s (t) \equiv \sqrt{\frac{z''}{z} + \frac{a^2 H^2}{4}} .
\end{equation}
Therefore $\ell \equiv a/s$ defines a new length scale in inflationary
cosmology --- the `potential scale'; it is $\ell$, not the horizon
scale $H^{-1}$ as is commonly stated in the literature, that controls
the form of the mode functions \cite{Martin:2003bt}. When a mode is
inside the potential scale it has the oscillatory behaviour of
eq.(\ref{UII}) and when it is outside it has the exponential behaviour
of eq.(\ref{3}). In de Sitter space we have $\ell=2/3H^{-1}$ and so
the two scales are nearly equal for slow-roll inflation. However, they
behave quite differently during the phase transition, with $\ell$
undergoing oscillations.

The oscillations in the curvature perturbation spectrum occur on
length scales which cross the `potential scale' as it is
oscillating. The maxima of the oscillations in $\mathcal{P_R}$
correspond to scales which cross at the minima of the potential
scale. These modes begin their exponential growth early and so have
increased by a larger amount at $x_{\rm f}$. Similarly modes which
cross at maxima of $\ell$ give rise to minima of $\mathcal{P_R}$. The
amplitude of the oscillations of the spectrum decreases with
wavenumber because the oscillations of $z''/z$ are damped.

\begin{figure}[tbh]
\includegraphics{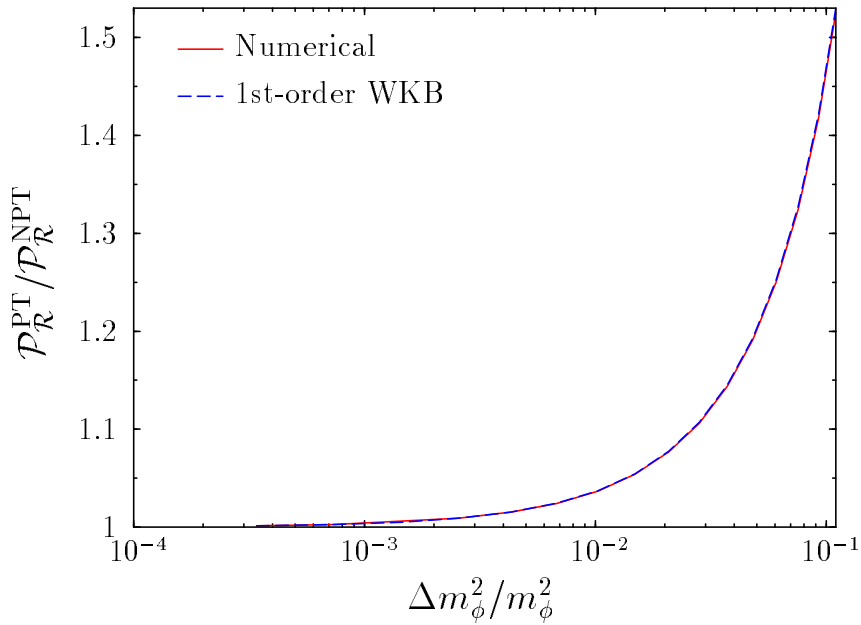}
\caption{ Comparison of the numerical result for the factor by which
 $\mathcal{P}_{\mathcal{R}}$ increases due to the phase transition with
 the 1st-order WKB expression, taking $k = 10\; \rm {h\; Mpc^{-1}}.$}.
\label{step} 
\end{figure}

Modes which cross the potential scale well before the phase transition
are growing according to $u_k \propto z$ by the time of the phase
transition, and so $\mathcal{P_R}$ is unchanged on large scales.
Modes which cross the potential scale well after the phase transition
are insensitive to the variation of $z''/z$ at the time of the phase
transition, because $k^2\gg z''/z$ then. However, they cross outside
the potential scale slightly later than would be the case if there
were no phase transition, because $\ell$ has increased from
$\left(\frac{9H^2}{4} + 6c_3\phi\right)^{-1/2}$ to
$\left(\frac{9H^2}{4}+6c_3\phi-\lambda\Sigma^2\right)^{-1/2}$.
Therefore for fixed $x_{\rm f}$ the modes will have undergone less
growth.  On the other hand, the introduction of the phase transition
causes an even larger decrease in the value of $\dot{\phi_{\rm f}}$
and so $\mathcal{P_R}$ increases on small scales. Consequently there is a
step in the spectrum. The factor by which the phase transition causes the 
spectrum to increase is given using the 1st-order WKB approximation as
\begin{eqnarray}
\frac{{\mathcal{P}_{\mathcal{R}}^{\mathrm{PT}}}(k)}
{\mathcal{P}_{\mathcal{R}}^{\mathrm{\mathrm{NPT}}}(k)} 
& = & 
\left[\frac{3c_3\left(\phi_{f}^{\mathrm{NPT}}\right)^2}{\left(3c_3\phi_{f}^{\mathrm{PT}}-\lambda\Sigma^{2}
\right)\phi_{f}^{\mathrm{PT}}}\right]^{2}\left[\frac{Q^{\mathrm{NPT}}\left(x_{f}\right)}{Q^{\mathrm{PT}}\left(x_{f}\right)}
\right]^{1/2} \nonumber \\
 &  & \times\exp\left\{ 2\int_{x_{*}^{\mathrm{PT}}}^
 {x_{f}}\left[-Q^{\mathrm{PT}}\left(y\right)\right]^{1/2}
 \mathrm{d}y-2\int_{x_{*}^{\mathrm{NPT}}}^{x_{f}}\left[-Q^{
 \mathrm{NPT}}\left(y\right)\right]^{1/2}\mathrm{d}y\right\}. 
\label{stepeq}
\end{eqnarray}
Here the superscripts `PT' and `NPT' denote quantities evaluated with
and without the phase transition, and we have used the slow-roll
expression $\dot{\phi}_{\mathrm{f}}\simeq -\frac{1}{3H}\frac{\partial
V}{\partial\phi}$ which is applicable at late times. The WKB
expression (\ref{stepeq}) is surprisingly accurate as can be seen from
Fig.\,\ref{step}.

\begin{figure}[tbh]
\includegraphics{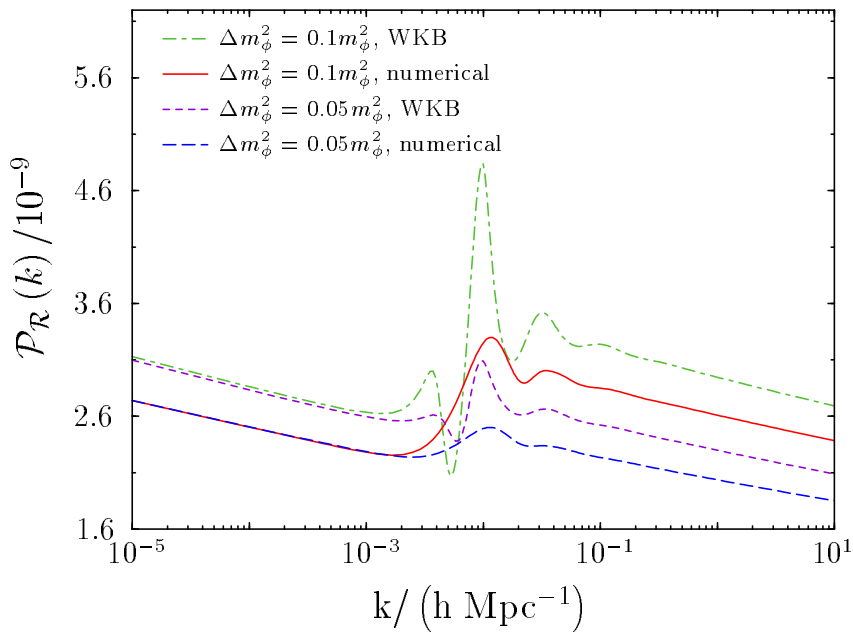}
\caption{ Comparison of the 2nd-order WKB approximation with the
  numerically calculated exact spectrum, for two different values of the
  change in the inflaton mass.}
\label{nswkb2} 
\end{figure}

\begin{figure}[tbh]
\includegraphics{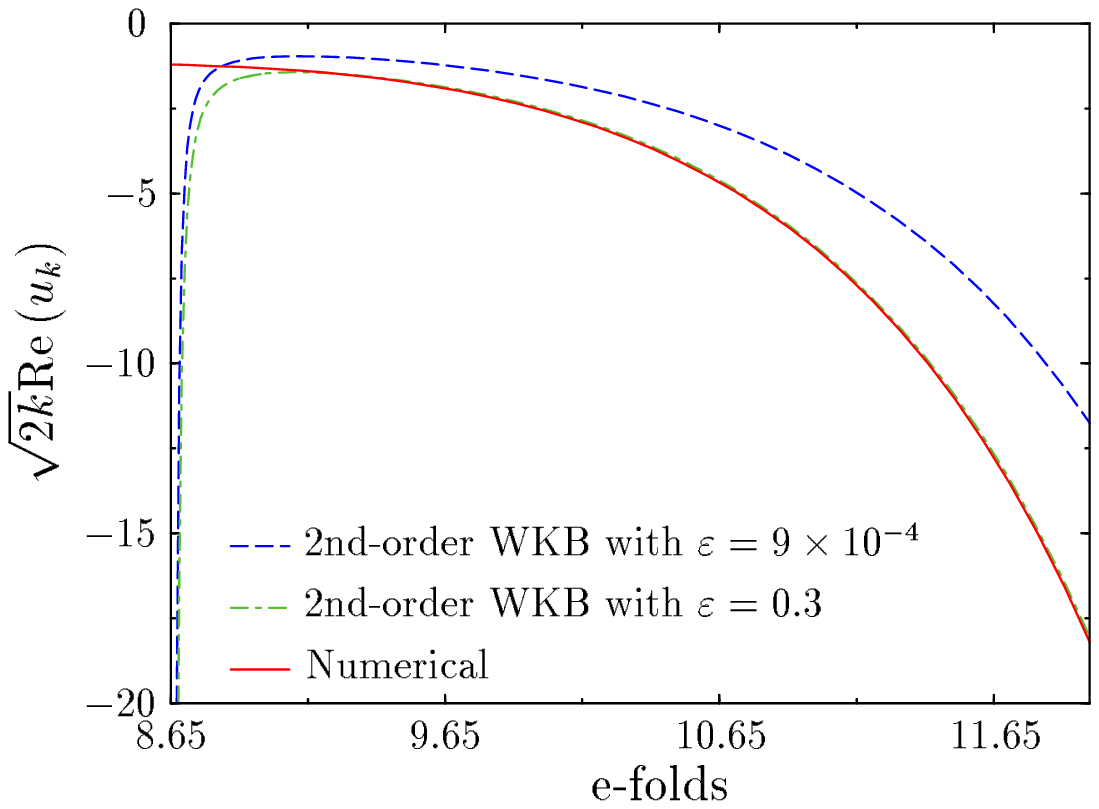}
\caption{ The effect on the 2nd-order WKB solution of varying $\varepsilon$.}
\label{varep} 
\end{figure}

\begin{figure}[tbh]
\includegraphics{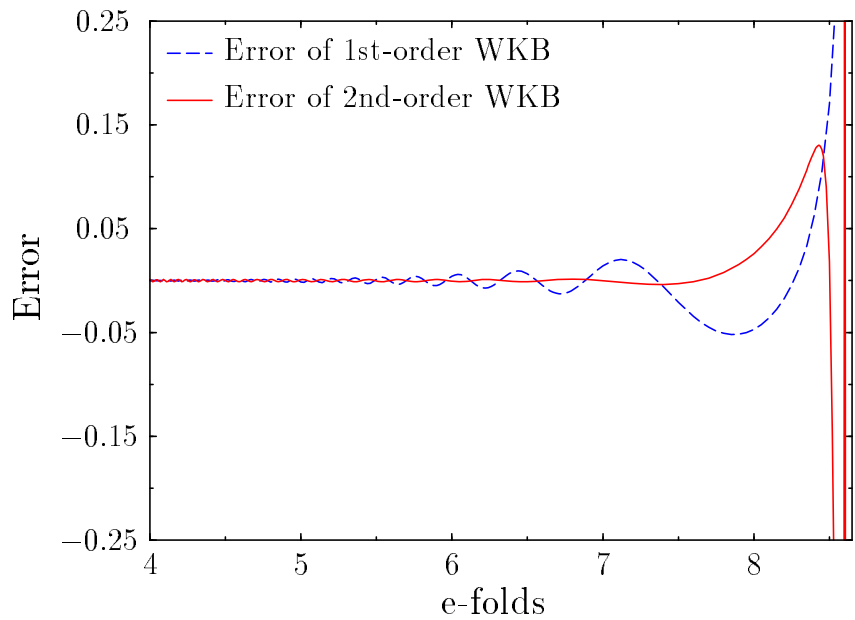}
\caption{ The error in $\sqrt{2k}\rm {Re}\left(u_{k}\right)$ for the
 1st- and 2nd-order WKB solutions.}
\label{erf} 
\end{figure}

As shown in Fig.\,\ref{nswkb2}, the spectrum calculated using the
2nd-order WKB approximation has an extra dip (at
$k\simeq5\times10^{-3}\;\mathrm{h\; Mpc^{-1}}$ in this example) and is
too large by $\sim10\%$ at small and large scales. The 2nd-order
solution is also weakly dependent on $\varepsilon$ as seen in
Fig.\,\ref{varep}; we noted earlier that this is due to the incomplete
cancellation between $\exp\left(\frac{\beta}{12\alpha^{3/2}}\varepsilon
^{-1/2}\right)$ and $\exp\left[\int_{x_*+\varepsilon}^{x_{\rm
f}}\frac{Q''}{48\left(-Q\right)^{3/2}}{\rm d}y\right]$ in
eq.(\ref{wkb2eq}). Overall the 2nd-order WKB approximation is in fact less
accurate than the 1st-order approximation because the asymptotic
matching is less successful. As can be seen from Fig.\,\ref{erf}, the
2nd-order solution is indeed more accurate that the 1st-order one away
from the turning point, but is less accurate close to $x_*$ where it
is more divergent. Hence the 2nd-order solution is less accurate in
region II, consequently eq.(\ref{will}) which is matched to it there
is less accurate than eq.(\ref{will2}) which matches the 1st-order
solution. The overlap between regions I and II, and between regions II
and III is reduced, so we have the seemingly paradoxical result that
the 1st-order approximation is superior to the 2nd-order one.

\section{Conclusions}

It is usually assumed that the curvature spectrum from primordial
inflation is scale-free, especially when fitting cosmological models
to data on CMB anisotropies and large-scale structure. This is indeed
the case for simple models of the inflaton where its potential is
assumed to be dominated by a (fine-tuned) single polynomial term in
the region where the observed fluctuations are generated. However it
must be emphasised that these are `toy' models and the actual physics
behind inflation is likely to be more complicated 
\cite{Quevedo:2002xw,Kachru:2003sx,Burgess:2004jk}. In particular when
one attempts to construct realistic models of inflation in the context
of supergravity or string theory, more than one field is typically
involved and their mutual interactions are unlikely to yield the
usually assumed smooth potential . Scalar fields other than the
inflaton can exhibit non-trivial dynamics during inflation and affect
the slow-roll evolution of the inflaton. In particular, `flat
direction' fields, which are generic in such models, may undergo
symmetry-breaking phase transitions and, by virtue of their
gravitational coupling to the inflaton, induce sudden changes in
its mass, resulting in multiple bursts of slow-roll inflation
punctuated by such phase transitions \cite{Adams:1997de}.

We have calculated analytically the effect of such sudden changes in
the inflaton mass on the curvature perturbation, using a WKB method
\cite{Martin:2002vn} to solve the governing Klein-Gordon equation
without recourse to any slow-roll approximations. Although the
analytic calculation does not yet provide an exact match to the exact
numerical solution, it does capture the characteristic oscillations
that are generated in the spectrum \cite{Starobinsky:ts,Adams:2001vc}
and provides considerable insight into the physics behind this
phenomenon. There may be scope for developing more accurate
calculational techniques in this context
\cite{Habib:2002yi,Habib:2004kc}. It should be of course borne in mind
that caution is required in using the standard calculational framework
for the inflationary metric perturbation for the case of a
time-dependent effective potential
\cite{Cooper:1996ii,Boyanovsky:1997xt}. Nevertheless when the scalar
fields involved are weakly coupled and the time scales involved exceed
the Hubble time, the usual results from perturbation theory ought to
be valid.

Given that the precision WMAP data does show preliminary evidence for
such oscillations \cite{Peiris:2003ff}, we are now investigating what
the data imply for the physics of the multiple inflation model and
indeed whether the estimation of cosmological parameters from the data
is significantly affected when the primordial spectrum is not
scale-free \cite{paper2}.

\begin{acknowledgments}

We are grateful to Jenni Adams, David Lyth, Graham Ross and Alexei
Starobinsky for discussions and to the anonymous Referee for a
critical and helpful report.

\end{acknowledgments}


\begin{thebibliography}{99}

%\cite{Spergel:2003cb}
\bibitem{Spergel:2003cb}
D.~N.~Spergel {\it et al.},
 %``First Year Wilkinson Microwave Anisotropy Probe (WMAP) Observations:
%Determination of Cosmological Parameters,''
Astrophys.\ J.\ Suppl.\  {\bf 148}, 175 (2003)
[arXiv:astro-ph/0302209].
%%CITATION = ASTRO-PH 0302209;%%

%\cite{Bridle:2003sa}
\bibitem{Bridle:2003sa}
S.~L.~Bridle, A.~M.~Lewis, J.~Weller and G.~Efstathiou,
%``Reconstructing the primordial power spectrum,''
Mon.\ Not.\ Roy.\ Astron.\ Soc.\  {\bf 342}, L72 (2003)
[arXiv:astro-ph/0302306].
%%CITATION = ASTRO-PH 0302306;%%

%\cite{Contaldi:2003zv}
\bibitem{Contaldi:2003zv}
C.~R.~Contaldi, M.~Peloso, L.~Kofman and A.~Linde,
%``Suppressing the lower Multipoles in the CMB Anisotropies,''
JCAP {\bf 0307}, 002 (2003)
[arXiv:astro-ph/0303636].
%%CITATION = ASTRO-PH 0303636;%%

%\cite{Blanchard:2003du}
\bibitem{Blanchard:2003du}
A.~Blanchard, M.~Douspis, M.~Rowan-Robinson and S.~Sarkar,
%``An alternative to the cosmological 'concordance model',''
Astron.\ Astrophys.\  {\bf 412}, 35 (2003)
[arXiv:astro-ph/0304237].
%%CITATION = ASTRO-PH 0304237;%%

%\cite{Cline:2003ve}
\bibitem{Cline:2003ve}
J.~M.~Cline, P.~Crotty and J.~Lesgourgues,
%``Does the small CMB quadrupole moment suggest new physics?,''
JCAP {\bf 0309}, 010 (2003)
[arXiv:astro-ph/0304558].
%%CITATION = ASTRO-PH 0304558;%%

%\cite{Feng:2003zu}
\bibitem{Feng:2003zu}
B.~Feng and X.~Zhang,
%``Double inflation and the low CMB quadrupole,''
Phys.\ Lett.\ B {\bf 570}, 145 (2003)
[arXiv:astro-ph/0305020].
%%CITATION = ASTRO-PH 0305020;%%

%\cite{Kawasaki:2003dd}
\bibitem{Kawasaki:2003dd}
M.~Kawasaki and F.~Takahashi,
%``Inflation model with lower multipoles of the CMB suppressed,''
Phys.\ Lett.\ B {\bf 570}, 151 (2003)
[arXiv:hep-ph/0305319].
%%CITATION = HEP-PH 0305319;%%

%\cite{Bastero-Gil:2003bv}
\bibitem{Bastero-Gil:2003bv}
M.~Bastero-Gil, K.~Freese and L.~Mersini-Houghton,
 %``What can WMAP tell us about the very early universe? New physics as an
%explanation of suppressed large scale power and running spectral index,''
Phys.\ Rev.\ D {\bf 68}, 123514 (2003)
[arXiv:hep-ph/0306289].
%%CITATION = HEP-PH 0306289;%%

%\cite{Kaloper:2003nv}
\bibitem{Kaloper:2003nv}
N.~Kaloper and M.~Kaplinghat,
%``Primeval corrections to the CMB anisotropies,''
Phys.\ Rev.\ D {\bf 68}, 123522 (2003)
[arXiv:hep-th/0307016].
%%CITATION = HEP-TH 0307016;%%

%\cite{Tsujikawa:2003gh}
\bibitem{Tsujikawa:2003gh}
S.~Tsujikawa, R.~Maartens and R.~Brandenberger,
%``Non-commutative inflation and the CMB,''
Phys.\ Lett.\ B {\bf 574}, 141 (2003)
[arXiv:astro-ph/0308169].
%%CITATION = ASTRO-PH 0308169;%%

%\cite{Moroi:2003pq}
\bibitem{Moroi:2003pq}
T.~Moroi and T.~Takahashi,
 %``Correlated isocurvature fluctuation in quintessence and suppressed CMB
%anisotropies at low multipoles,''
Phys.\ Rev.\ Lett.\  {\bf 92}, 091301 (2004)
[arXiv:astro-ph/0308208].
%%CITATION = ASTRO-PH 0308208;%%

%\cite{Piao:2003wp}
\bibitem{Piao:2003wp}
Y.~S.~Piao, B.~Feng and X.~Zhang,
%``Suppressing CMB quadrupole with a bounce from contracting phase 
%to inflation,''
Phys.\ Rev.\ D {\bf 69}, 103520 (2004)
[arXiv:hep-th/0310206].
%%CITATION = HEP-TH 0310206;%%

%\cite{Gaztanaga:2003ub}
\bibitem{Gaztanaga:2003ub}
E.~Gaztanaga, J.~Wagg, T.~Multamaki, A.~Montana and D.~H.~Hughes,
%``2-point anisotropies in WMAP and the Cosmic Quadrupole,''
Mon.\ Not.\ Roy.\ Astron.\ Soc.\  {\bf 346}, 47 (2003)
[arXiv:astro-ph/0304178].
%%CITATION = ASTRO-PH 0304178;%%

%\cite{Efstathiou:2003wr}
\bibitem{Efstathiou:2003wr}
G.~Efstathiou,
%``The Statistical Significance of the Low CMB Mulitipoles,''
Mon.\ Not.\ Roy.\ Astron.\ Soc.\  {\bf 346}, L26 (2003)
[arXiv:astro-ph/0306431].
%%CITATION = ASTRO-PH 0306431;%%

%\cite{deOliveira-Costa:2003pu}
\bibitem{deOliveira-Costa:2003pu}
A.~de Oliveira-Costa, M.~Tegmark, M.~Zaldarriaga and A.~Hamilton,
%``The significance of the largest scale CMB fluctuations in WMAP,''
Phys.\ Rev.\ D {\bf 69}, 063516 (2004)
[arXiv:astro-ph/0307282].
%%CITATION = ASTRO-PH 0307282;%%

%\bibitem{Schwarz:2004gk}
\bibitem{Schwarz:2004gk}
D.~J.~Schwarz, G.~D.~Starkman, D.~Huterer and C.~J.~Copi,
%``Is the low-l microwave background cosmic?,''
arXiv:astro-ph/0403353.
%%CITATION = ASTRO-PH 0403353;%%

%\cite{Eriksen:2003db}
\bibitem{Eriksen:2003db}
H.~K.~Eriksen, F.~K.~Hansen, A.~J.~Banday, K.~M.~Gorski and P.~B.~Lilje,
%``Asymmetries in the CMB anisotropy field,''
Astrophys.\ J.\ {\bf 605}, 14 (2004)
[arXiv:astro-ph/0307507].
%%CITATION = ASTRO-PH 0307507;%%

%\cite{Peiris:2003ff}
\bibitem{Peiris:2003ff}
H.~V.~Peiris {\it et al.},
 %``First year Wilkinson Microwave Anisotropy Probe (WMAP) observations:
%Implications for inflation,''
Astrophys.\ J.\ Suppl.\  {\bf 148}, 213 (2003)
[arXiv:astro-ph/0302225].
%%CITATION = ASTRO-PH 0302225;%%

%\cite{Adams:1997de}
\bibitem{Adams:1997de}
J.~A.~Adams, G.~G.~Ross and S.~Sarkar,
%``Multiple inflation,''
Nucl.\ Phys.\ B {\bf 503}, 405 (1997)
[arXiv:hep-ph/9704286].
%%CITATION = HEP-PH 9704286;%%

%\cite{Adams:2001vc}
\bibitem{Adams:2001vc}
J.~Adams, B.~Cresswell and R.~Easther,
%``Inflationary perturbations from a potential with a step,''
Phys.\ Rev.\ D {\bf 64}, 123514 (2001)
[arXiv:astro-ph/0102236].
%%CITATION = ASTRO-PH 0102236;%%

%\cite{Martin:2002vn}
\bibitem{Martin:2002vn}
J.~Martin and D.~J.~Schwarz,
%``WKB approximation for inflationary cosmological perturbations,''
Phys.\ Rev.\ D {\bf 67}, 083512 (2003)
[arXiv:astro-ph/0210090].
%%CITATION = ASTRO-PH 0210090;%%

\bibitem{paper2}
M. Douspis, P. Hunt and S. Sarkar,
in preparation.

%\cite{Wang:2002hf}
\bibitem{Wang:2002hf}
X.~Wang, B.~Feng and M.~Li,
 %``Natural inflation, Planck scale physics and oscillating primordial
%spectrum,''
arXiv:astro-ph/0209242.
%%CITATION = ASTRO-PH 0209242;%%

%\cite{Burgess:2002ub}
\bibitem{Burgess:2002ub}
C.~P.~Burgess, J.~M.~Cline, F.~Lemieux and R.~Holman,
%``Are inflationary predictions sensitive to very high energy physics?,''
JHEP {\bf 0302}, 048 (2003)
[arXiv:hep-th/0210233].
%%CITATION = HEP-TH 0210233;%%

%\cite{Martin:2003sg}
\bibitem{Martin:2003sg}
J.~Martin and C.~Ringeval,
%``Superimposed Oscillations in the WMAP Data?,''
Phys.\ Rev.\ D {\bf 69}, 083515 (2004)
[arXiv:astro-ph/0310382].
%%CITATION = ASTRO-PH 0310382;%%

%\cite{Okamoto:2003wk}
\bibitem{Okamoto:2003wk}
T.~Okamoto and E.~A.~Lim,
%``Constraining Cut-off Physics in the Cosmic Microwave Background,''
Phys.\ Rev.\ D {\bf 69}, 083519 (2004)
[arXiv:astro-ph/0312284].
%%CITATION = ASTRO-PH 0312284;%%

%\cite{Martin:2004iv}
\bibitem{Martin:2004iv}
J.~Martin and C.~Ringeval,
%``Addendum to ``Superimposed Oscillations in the WMAP Data?'',''
Phys.\ Rev.\ D {\bf 69}, 127303 (2004)
[arXiv:astro-ph/0402609].
%%CITATION = ASTRO-PH 0402609;%%

%\cite{Chung:1999ve}
\bibitem{Chung:1999ve}
D.~J.~H.~Chung, E.~W.~Kolb, A.~Riotto and I.~I.~Tkachev,
%``Probing Planckian physics: Resonant production of particles during 
%inflation and features in the primordial power spectrum,''
Phys.\ Rev.\ D {\bf 62} (2000) 043508
[arXiv:hep-ph/9910437].  
%%CITATION = HEP-PH 9910437;%%

%\cite{Elgaroy:2003hp}
\bibitem{Elgaroy:2003hp}
O.~Elgaroy, S.~Hannestad and T.~Haugboelle,
%``Observational constraints on particle production during inflation,''
JCAP {\bf 0309} (2003) 008
[arXiv:astro-ph/0306229].
%%CITATION = ASTRO-PH 0306229;%%

%\cite{Kawasaki:2004pi}
\bibitem{Kawasaki:2004pi}
M.~Kawasaki, F.~Takahashi and T.~Takahashi,
%``Making waves on CMB power spectrum and inflaton dynamics,''
arXiv:astro-ph/0407631.
%%CITATION = ASTRO-PH 0407631;%%

%\cite{Mukherjee:2003ag}
\bibitem{Mukherjee:2003ag}
P.~Mukherjee and Y.~Wang,
 %``Model-Independent Reconstruction of the Primordial Power Spectrum from WMAP
%Data,''
Astrophys.\ J.\  {\bf 599}, 1 (2003)
[arXiv:astro-ph/0303211].
%%CITATION = ASTRO-PH 0303211;%%

%\cite{Hannestad:2003zs}
\bibitem{Hannestad:2003zs}
S.~Hannestad,
%``Reconstructing the primordial power spectrum - a new algorithm,''
JCAP {\bf 0404}, 002 (2004)
[arXiv:astro-ph/0311491].
%%CITATION = ASTRO-PH 0311491;%%

%\cite{Shafieloo:2003gf}
\bibitem{Shafieloo:2003gf}
A.~Shafieloo and T.~Souradeep,
%``Primordial power spectrum from WMAP,''
Phys.\ Rev.\ D {\bf 70}, 043523 (2004)
[arXiv:astro-ph/0312174].
%%CITATION = ASTRO-PH 0312174;%%

%\cite{Kogo:2003yb}
\bibitem{Kogo:2003yb}
N.~Kogo, M.~Matsumiya, M.~Sasaki and J.~Yokoyama,
%``Reconstructing the primordial spectrum from WMAP data by the cosmic
%inversion method,''
Astrophys.\ J.\ {\bf 607}, 32 (2004)
[arXiv:astro-ph/0309662].
%%CITATION = ASTRO-PH 0309662;%%

%\cite{Tocchini-Valentini:2004ht}
\bibitem{Tocchini-Valentini:2004ht}
D.~Tocchini-Valentini, M.~Douspis and J.~Silk,
%``Are there features in the primordial power spectrum?,''
arXiv:astro-ph/0402583.
%%CITATION = ASTRO-PH 0402583;%%

%\cite{Nilles:1983ge}
\bibitem{Nilles:1983ge}
H.~P.~Nilles,
%``Supersymmetry, Supergravity And Particle Physics,''
Phys.\ Rept.\  {\bf 110}, 1 (1984).
%%CITATION = PRPLC,110,1;%%

%\cite{Chung:2003fi}
\bibitem{Chung:2003fi}
D.~J.~H.~Chung, L.~L.~Everett, G.~L.~Kane, S.~F.~King, J.~Lykken and L.~T.~Wang,
%``The soft supersymmetry-breaking Lagrangian: Theory and applications,''
arXiv:hep-ph/0312378.
%%CITATION = HEP-PH 0312378;%%

%\cite{Enqvist:2003gh}
\bibitem{Enqvist:2003gh}
K.~Enqvist and A.~Mazumdar,
%``Cosmological consequences of MSSM flat directions,''
Phys.\ Rept.\  {\bf 380}, 99 (2003).
[arXiv:hep-ph/0209244].
%%CITATION = HEP-PH 0209244;%%

%\cite{Coughlan:yk}
\bibitem{Coughlan:yk}
G.~D.~Coughlan, R.~Holman, P.~Ramond and G.~G.~Ross,
%``Supersymmetry And The Entropy Crisis,''
Phys.\ Lett.\ B {\bf 140}, 44 (1984).
%%CITATION = PHLTA,B140,44;%%

%\cite{Copeland:1994vg}
\bibitem{Copeland:1994vg}
E.~J.~Copeland, A.~R.~Liddle, D.~H.~Lyth, E.~D.~Stewart and D.~Wands,
%``False vacuum inflation with Einstein gravity,''
Phys.\ Rev.\ D {\bf 49}, 6410 (1994)
[arXiv:astro-ph/9401011].
%%CITATION = ASTRO-PH 9401011;%%

%\cite{Dine:1995uk}
\bibitem{Dine:1995uk}
M.~Dine, L.~Randall and S.~Thomas,
%``Supersymmetry breaking in the early universe,''
Phys.\ Rev.\ Lett.\  {\bf 75}, 398 (1995)
[arXiv:hep-ph/9503303].
%%CITATION = HEP-PH 9503303;%%

%\cite{Lyth:1998xn}
\bibitem{Lyth:1998xn}
D.~H.~Lyth and A.~Riotto,
 %``Particle physics models of inflation and the cosmological density
%perturbation,''
Phys.\ Rept.\  {\bf 314}, 1 (1999)
[arXiv:hep-ph/9807278].
%%CITATION = HEP-PH 9807278;%%

%\cite{Yamamoto:1985rd}
\bibitem{Yamamoto:1985rd}
K.~Yamamoto,
 %``Phase Transition Associated With Intermediate Gauge Symmetry Breaking In
%Superstring Models,''
Phys.\ Lett.\ B {\bf 168}, 341 (1986).
%%CITATION = PHLTA,B168,341;%%

%\cite{Barreiro:1996dx}
\bibitem{Barreiro:1996dx}
T.~Barreiro, E.~J.~Copeland, D.~H.~Lyth and T.~Prokopec,
 %``Some aspects of thermal inflation: the finite temperature potential and
%topological defects,''
Phys.\ Rev.\ D {\bf 54}, 1379 (1996)
[arXiv:hep-ph/9602263].
%%CITATION = HEP-PH 9602263;%%

%\cite{Ross:1995dq}
\bibitem{Ross:1995dq}
G.~G.~Ross and S.~Sarkar,
%``Successful supersymmetric inflation,''
Nucl.\ Phys.\ B {\bf 461}, 597 (1996)
[arXiv:hep-ph/9506283].
%%CITATION = HEP-PH 9506283;%%

%\cite{Adams:1996yd}
\bibitem{Adams:1996yd}
J.~A.~Adams, G.~G.~Ross and S.~Sarkar,
%``Natural supergravity inflation,''
Phys.\ Lett.\ B {\bf 391}, 271 (1997)
[arXiv:hep-ph/9608336].
%%CITATION = HEP-PH 9608336;%%

%\cite{Garcia-Bellido:1995qq}
\bibitem{Garcia-Bellido:1995qq}
J.~Garcia-Bellido and D.~Wands,
%``Metric perturbations in two-field inflation,''
Phys.\ Rev.\ D {\bf 53}, 5437 (1996)
[arXiv:astro-ph/9511029].
%%CITATION = ASTRO-PH 9511029;%%

%\cite{Sasaki:1995aw}
\bibitem{Sasaki:1995aw}
M.~Sasaki and E.~D.~Stewart,
 %``A General analytic formula for the spectral index of the density
%perturbations produced during inflation,''
Prog.\ Theor.\ Phys.\  {\bf 95}, 71 (1996)
[arXiv:astro-ph/9507001].
%%CITATION = ASTRO-PH 9507001;%%

%\cite{Starobinsky:ts}
\bibitem{Starobinsky:ts}
A.~A.~Starobinsky,
 %``Spectrum Of Adiabatic Perturbations In The Universe When There Are
%Singularities In The Inflation Potential,''
JETP Lett.\  {\bf 55}, 489 (1992)
[Pisma Zh.\ Eksp.\ Teor.\ Fiz.\  {\bf 55}, 477 (1992)].
%%CITATION = JTPLA,55,489;%%

%\cite{Leach:2001zf}
\bibitem{Leach:2001zf}
S.~M.~Leach, M.~Sasaki, D.~Wands and A.~R.~Liddle,
%``Enhancement of superhorizon scale inflationary curvature perturbations,''
Phys.\ Rev.\ D {\bf 64}, 023512 (2001)
[arXiv:astro-ph/0101406].
%%CITATION = ASTRO-PH 0101406;%%

%\cite{Stewart:1993bc}
\bibitem{Stewart:1993bc}
E.~D.~Stewart and D.~H.~Lyth,
 %``A More accurate analytic calculation of the spectrum of cosmological
%perturbations produced during inflation,''
Phys.\ Lett.\ B {\bf 302}, 171 (1993)
[arXiv:gr-qc/9302019].
%%CITATION = GR-QC 9302019;%%

%\cite{Martin:2003bt}
\bibitem{Martin:2003bt}
J.~Martin,
%``Inflation and precision cosmology,''
arXiv:astro-ph/0312492.
%%CITATION = ASTRO-PH 0312492;%%

%\cite{Mukhanov:1990me}
\bibitem{Mukhanov:1990me}
V.~F.~Mukhanov, H.~A.~Feldman and R.~H.~Brandenberger,
 %``Theory Of Cosmological Perturbations. Part 1. Classical Perturbations. Part
%2. Quantum Theory Of Perturbations. Part 3. Extensions,''
Phys.\ Rept.\ {\bf 215}, 203 (1992).
%%CITATION = PRPLC,215,203;%%

%\cite{Malik:2001rm}
\bibitem{Malik:2001rm}
K.~A.~Malik,
%``Cosmological perturbations in an inflationary universe,''
arXiv:astro-ph/0101563.
%%CITATION = ASTRO-PH 0101563;%%

%\cite{Riotto:2002yw}
\bibitem{Riotto:2002yw}
A.~Riotto,
%``Inflation and the theory of cosmological perturbations,''
arXiv:hep-ph/0210162.
%%CITATION = HEP-PH 0210162;%%

%\cite{Habib:2002yi}
\bibitem{Habib:2002yi}
S.~Habib, K.~Heitmann, G.~Jungman and C.~Molina-Paris,
%``The inflationary perturbation spectrum,''
Phys.\ Rev.\ Lett.\  {\bf 89}, 281301 (2002)
[arXiv:astro-ph/0208443].
%%CITATION = ASTRO-PH 0208443;%%

%\cite{Habib:2004kc}
\bibitem{Habib:2004kc}
S.~Habib, A.~Heinen, K.~Heitmann, G.~Jungman and C.~Molina-Paris,
%``Characterizing inflationary perturbations: The uniform approximation,''
arXiv:astro-ph/0406134.
%%CITATION = ASTRO-PH 0406134;%%

\bibitem{wkb}
C.~M.~Bender and S.~A.~Orszag,
``Advanced mathematical Methods for Scientists and Engineers'',
(McGraw-Hill, 1978).

%\cite{Quevedo:2002xw}
\bibitem{Quevedo:2002xw}
F.~Quevedo,
%``Lectures on string / brane cosmology,''
Class.\ Quant.\ Grav.\  {\bf 19}, 5721 (2002)
[arXiv:hep-th/0210292].
%%CITATION = HEP-TH 0210292;%%

%\cite{Kachru:2003sx}
\bibitem{Kachru:2003sx}
S.~Kachru, R.~Kallosh, A.~Linde, J.~Maldacena, L.~McAllister and S.~P.~Trivedi,
%``Towards inflation in string theory,''
JCAP {\bf 0310}, 013 (2003)
[arXiv:hep-th/0308055].
%%CITATION = HEP-TH 0308055;%%

%\cite{Burgess:2004jk}
\bibitem{Burgess:2004jk}
C.~P.~Burgess,
%``Inflatable string theory?,''
arXiv:hep-th/0408037.
%%CITATION = HEP-TH 0408037;%%

%\cite{Cooper:1996ii}
\bibitem{Cooper:1996ii}
F.~Cooper, S.~Habib, Y.~Kluger and E.~Mottola,
%``Nonequilibrium dynamics of symmetry breaking in lambda Phi**4 field
%theory,''
Phys.\ Rev.\ D {\bf 55}, 6471 (1997)
[arXiv:hep-ph/9610345].
%%CITATION = HEP-PH 9610345;%%

%\cite{Boyanovsky:1997xt}
\bibitem{Boyanovsky:1997xt}
D.~Boyanovsky, D.~Cormier, H.~J.~de Vega, R.~Holman and S.~P.~Kumar,
%``Non-perturbative quantum dynamics of a new inflation model,''
Phys.\ Rev.\ D {\bf 57}, 2166 (1998)
[arXiv:hep-ph/9709232].
%%CITATION = HEP-PH 9709232;%%

\end{thebibliography}
\end{document}